\documentclass[lettersize,journal]{IEEEtran}
\usepackage{amsmath,amsfonts}
\usepackage{multirow}
\usepackage{algorithmic}
\usepackage{algorithm}
\usepackage{array}
\usepackage[caption=false,font=normalsize,labelfont=sf,textfont=sf]{subfig}
\usepackage{textcomp}
\usepackage{stfloats}
\usepackage{url}
\usepackage{verbatim}
\usepackage{graphicx}
\usepackage{cite}
\usepackage{colortbl,pifont}
\definecolor{color1}{rgb}{0.984,0.894,0.835}
\definecolor{color2}{rgb}{0.886,0.937,0.851}
\begin{document}

\title{Novel Phase detector measurement procedure\\ using Quasi‐Synchronized RF Generator}

\author{V\'ictor Ara\~na Pulido, ~\IEEEmembership{Member,~IEEE,}
Francisco Cabrera-Almeida, ~\IEEEmembership{Member,~IEEE,}
Pedro Quintana-Morales,
Eduardo Mendieta-Otero
\thanks{The authors are with the Institute for Technological Development and Innovation in Communications (IDeTIC),
Department of Signals and Communication, University of Las Palmas de Gran Canaria (ULPGC), 35017 Las Palmas, Spain
(e-mail: victor.arana@ulpgc.es; francisco.cabrera@ulpgc.es; pedro.quintana@ulpgc.es; eduardo.mendieta@ulpgc.es).}
\thanks{This work was supported by the Spanish Government under Grant PID2020-116569RB-C32 Project.}
\thanks{Manuscript received April 19, 2021; revised August 16, 2021.}}

\markboth{Journal of \LaTeX\ Class Files,~Vol.~14, No.~8, August~2021}%
{Shell \MakeLowercase{\textit{et al.}}: A Sample Article Using IEEEtran.cls for IEEE Journals}


\maketitle

\begin{abstract}
This paper presents a new procedure for phase detector measurements that allows the use of generators that share a 10 MHz reference oscillator but do not synchronize in phase, in other words, quasi-synchronized RF generators. The objectives are taking advantage of the benefits of using two generators but recovering lower-cost generators that have worse synchronization performance and opening the door to the possibility of using a very simple control element based in Arduino Uno and cheaper instruments. The new procedure is characterized by continuously alternating calibration and measurement sequences to make up for the phase drift of quasi-synchronized generators and guarantee a maximum phase error specification ($\pm$1\textordmasculine\ in this paper). Data acquisition has been divided in two stages: measurement of detector curves without phase reference (in-phase and phase-shifted) and measurement of reference data. All the data is later combined to obtain correctly referenced in-phase detector curves. The technique can be reproduced with other equivalent instrumentation. The novel procedure that allows compensation for errors (amplitude, phase shift, mismatching, etc.) is detailed, and its relation to the required measurement accuracy is amply discussed. The proposed technique is applied to characterize a phase detector based on in-phase and phase-shifted multiplication from 3 to 8 GHz with 1 GHz step. Measurements have a final maximum error of $\pm$2\textordmasculine\ for both frequency and calibrated input power, according to the accuracy specifications of the VNA used to calibrate the signal distribution network, added to the $\pm$1\textordmasculine\ specified in this new procedure.
\end{abstract}


\section{Introduction}
\IEEEPARstart{P}{hase} detectors are used in antenna arrays, PLL (Phase-Locked Loop), interferometry, measurement of physical parameters, distance measurement, positioning systems, etc. A phase detector provides a voltage ($V_d$) that is proportional to the phase difference between the input signals ($\Delta\theta$). The characteristic curve of a phase detector derives from the output voltage versus the phase difference between input signals. In order to perform this measurement, it is necessary to have two signals with the same frequency (synchronized) and modify the phase difference between them. The higher the frequency, the harder it is for the circuit to generate these signals. Additionally, a small difference between the signal paths to the device generates an error in the measurement that grows as the frequency increases.

When the phase detectors are analog multipliers with a 90\textordmasculine\ phase shift in one of their paths and a sinusoidal response is assumed ($V_d = \sin \Delta\theta$), it is common to use input signals with zero phase shift and adjust the circuit until a null is achieved in the output voltage \cite{Pogorzelski2005}. A generator connected to a power divider and two identical wires are more than sufficient to make the measurement, which results in significant cost savings. However, this phase detector topology needs adding a VNA (Vector Network Analyzer) which is used to characterize the asymemetrical paths (amplifiers, hybrids, etc.) that carry the signals to the multiplier. Moreover, if the frequency increases, a multiplier can be added to each path, which makes the measurement system substantially more expensive when using commercial circuits \cite{Philippe2018,RSSMZ}.

When the phase detector response is more irregular and a higher accuracy is required \cite{Perez2021,Yang2014, Choi2019, Hua2016}, the entire curve must be plotted within the phase operating range, which requires having knowledge of the input phase shift at all times.

Some measurement systems use generators that are equipped with two or more RF outputs ($<$ 1 GHz) and phase shift circuits that can be easily calibrated \cite{Yang2014, Wang2017}. When the frequency increases, it is more common to find generators with a single RF output. In that case, a power divider is connected to RF output and a phase shifter is included in one of the paths \cite{Hua2016, Hua2011, Hua2009}. There are applications that only require measurements around a specific phase shift or where the authors have no access to more complex and expensive equipment. For example, in \cite{Yang2014}, measurements are taken from 4.5 GHz to 5.5 GHz using one generator and a hybrid circuit. Additionally, if the phase detector is sensitive to amplitude variation at the inputs, a variable gain stage is usually added to compensate the amplitudes of both paths \cite{Huang2022}. As references show, when a single generator is used, a VNA is necessary to characterize and calibrate the significant asymmetries between paths and be able to trace detector curves. The parameters that provide the input phase shift and amplitude to the DUT (Device Under Test) for each frequency, power, temperature, etc. must be controlled. Although the error of the measurement system is not indicated, it will at least depend on the accuracy of the VNA.

The use of two synchronized commercial generators facilitates measurement and makes the use of a VNA non mandatory. Phase shifts and amplitudes can be easily modified and the devices that carry the signals to the DUT inputs can be identical in each path (cables, switches, circulators, attenuators, etc.), which minimizes the imbalance. When the frequencies are low, it is enough to share a reference oscillator, usually of 10 MHz, and use the phase shift option provided by the generators. Nevertheless, these devices often include additional synthesis and frequency conversion stages that prevent real synchronization, resulting in phase shift drift. This problem becomes more critical as the output frequency of the generators increases. Therefore, it is also necessary to share LO (Local Oscilator) of these mixings \cite{Perez2021, 1GP67, Umpierrez2012} to improve synchronization. There is commercial equipment available that includes this option, that is capable of maintaining good synchronization over time up to 6 GHz \cite{1GP67, 5991-0038}. This enables the entire calibration process to be carried out first, followed by the measurement process. The cost of each generator with this option is about twice as much as one with a similar frequency range and that only shares the classic 10 MHz reference signal. These equipments can be used to characterize most phase detectors operating in these frequency bands, such as digital \cite{Rezaeian2020, Sofimowloodi2019} or FPGA (Field Programmable Gate Array) based \cite{Wang2017, Zhang2018, Vandenbussche2014}. In addition to these, there is a long list of applications that use the AD8302 phase detector \cite{AD8302} as a phase detector in the 0-180\textordmasculine\ range \cite{Lopez2020} or within $IQ$ configurations to achieve 0-360\textordmasculine\ \cite{Mlynek2017}. By sharing LO in each path, these detectors are incorporated in superheterodyne designs that raise the input frequency \cite{Mlynek2017, Podobedov1998, Varavin2016, Drozd2022}. However, the characterization of the phase detector has been done in IF ($<$ 133 MHz), avoiding the problem of generating the calibrated RF signals.

Finally, regarding the characterization of errors in all referenced measurement systems, only \cite{Perez2021} and \cite{Umpierrez2012} include a comprehensive investigation that guarantees an error of less than $\pm$0.1\textordmasculine\ up to 6 GHz for two synchronized commercial generators.

\begin{figure}[!t]
\centering
\includegraphics[width=4.0cm]{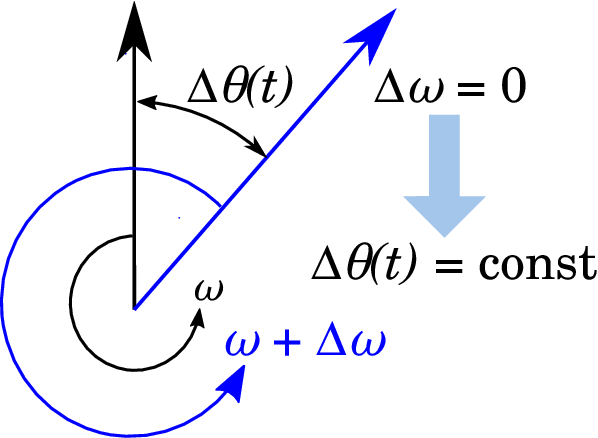}
\caption{Phasor representation of two signals of frequency $w$ and $w$ + $\Delta w$ ($w=2\cdot \pi\cdot f$).}
\label{fig:figura1}
\end{figure}

This article proposes a method for measuring phase detectors using two generators that can only share the 10 MHz reference, which implies a small difference between their output frequencies (quasi-synchronized) and thus a phase shift drift (Fig. \ref{fig:figura1}). The procedure is characterized by continuously alternating calibration and measurement sequences to guarantee a maximum phase error specification. The objective is to take advantage of the benefits of using two generators while recovering generators of lower cost and worse performance in terms of synchronization. In addition to the generators, the calibration and measurement system consists of two switches, a power combiner, a power detector (a spectrum analyzer in this article), an oscilloscope, and an Arduino Uno with ADC (Analog to Digital Converter) included. The signal distribution network errors are assessed to specify how they should be taken into account in the calibration and measurement process and to establish whether the employment of the VNA is required. In the frequency range between 3 GHz and 8 GHz, the method is used to obtain the curves of a 360\textordmasculine\ switched dual multiplier phase detector \cite{Perez2021}.

This paper is organized as follows. Section II describes the differences between synchronized and quasi-synchronized generators, clarifying certain aspects related to calibration in each case, as well as the sign convention for the phases. Section III presents the calibration and measurement system, which details the composition of devices and equipment. Section IV reviews some aspects of the DUT, the 360\textordmasculine\ phase detector base on switched dual multiplier \cite{Perez2021}, in order to determine the measures to be performed to characterize this type of devices \cite{Perez2017}. Additionally, the sign convention of the multiplier output signals is justified. Section V addresses aspects relating to the calibration of the signal distribution network and its effect on the semi-automated calibration and measurement procedure that is detailed in Section VI. Lastly, Section VII describes the procedure to reference the curves and determine the operating range of the DUT, attending to the phase deviation from the quadrature condition (90\textordmasculine\ phase shift).

\section{Synchronized and Quasi‐Synchronized generators: sign conventions \& calibration}
Commercial generators that are able to share the reference signal, usually of 10 MHz, typically have the possibility to modify the phase of each of them, that is, to modify the phase shift between both RF signals. When two generators are properly synchronized, the frequencies match and the phase difference between the signals remains constant ($\Delta w=0$ in Fig. \ref{fig:figura1}). This article uses Agilent models E8257D and E8257C. Model C (primary) provides the 10 MHz reference signal to model D (secondary). To allay concerns about the sign convention that will be used in this paper when modifying the phase of these generators, we proceed as follows. First, choose a low frequency that can be displayed on the oscilloscope ($<$ 20 MHz) and select the value 0\textordmasculine\ on the primary generator. Next, the phase of the secondary generator ($\theta_{S_{REF_0}}$) is modified until the oscilloscope signals are in phase ($\Delta\theta=0$).
If the oscilloscope trigger is synchronized with the secondary generator signal, a phase shift in the primary generator signal equal to the selected value in the primary generator phase is observed.
That means that if the secondary generator signal is $A \cdot \cos(wt)$, the primary signal is given by $A\cdot\cos(wt+\theta_M)$.

When the frequency increases, finding the value of the reference phase ($\theta_{S_{REF_0}}$) using the oscilloscope technique is unfeasible in most cases, whether due to technical constraints or due to the cost of instrumentation. In that case, the null technique \cite{1GP67, Umpierrez2012} is usually used to obtain the reference signal ($\theta_{S_{REF_{180}}}$). The signals of the generators are fed into a power combiner, the phase of the primary is set to 180\textordmasculine\ ($\theta_M =$ 180\textordmasculine) and the phase of the secondary ($\theta_{S_{REF_{180}}}$) is modified until a null is obtained at the output of the combiner. However, when there are amplitude and phase deviations between the generator signals, it is of greater interest to analyze the value of the null ($\theta_M =$ 180\textordmasculine) versus the maximum ($\theta_M =$ 0\textordmasculine).

\begin{figure}[!t]
\centering
\includegraphics[width=7.3cm]{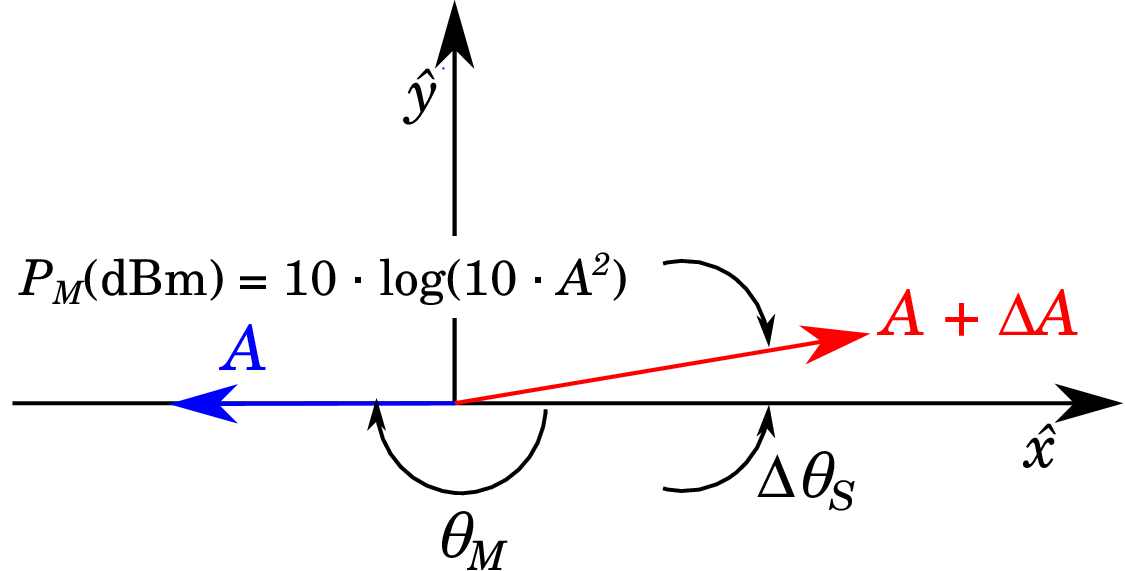}
\caption{Representation of the amplitude phasors at the input of the RF combiner. It shows the effect of amplitude ($\Delta A$) and phase ($\Delta\theta_S$) deviation on both the maximum power at the output ($\theta_M =$ 0\textordmasculine), and the minimum power at the output ($\theta_M =$ 180\textordmasculine).}
\label{fig:figura2}
\end{figure}

If modifying the phase of the primary generator $A\cdot\cos(wt+\theta_M)$ is preferred, the null at the output of the RF combiner occurs when $\theta_M =$ 180\textordmasculine\ and $\Delta \theta_S =$ 0\textordmasculine\ (Fig. \ref{fig:figura2}). If amplitude and phase deviations of the secondary generator signal equal to $\Delta A$ and $\Delta\theta_S$) are assumed, the output signal of RF combiner for $\theta_M =$ 0\textordmasculine\ and 180\textordmasculine, that correspond with maximum and minimun values, is given  by (Fig. \ref{fig:figura2}):
\begin{equation}
\small
 A \cos(\theta_M)\hat{x}+\left(A+\Delta A\right) \cos(\Delta\theta_S)\hat{x} + \left(A+\Delta A\right)\sin(\Delta\theta_S)\hat{y}
\end{equation}
The ratio between the minimum and maximum power is:
\begin{equation}
\small
\begin{split}
r\textrm{ (dB)} & = 10 \cdot \log (R) \\
 & = 10 \cdot \log\left(\frac{\Delta A^2+2A(A+\Delta A)(1-\cos(\Delta\theta_S))}{\Delta A^2 + 2A(A+\Delta A)(1+\cos(\Delta\theta_S))}\right)
\end{split}
\end{equation}

Table \ref{tab:table_i} shows some relevant values, when the power is $a$ dBm ($A = 10^{a/20 -1/2}$) and the amplitude ratio $(A+\Delta A)/A$ is  given by $s$ dB ($\Delta A = (10^{s/20} - 1)\cdot A$). If one assumes that amplitudes are identical ($s =$ 0 dB), an error of less than $\pm$1\textordmasculine\ is guaranteed for a drop in power of more than 41 dB measured in the upper SA, regardless of the input level (cases 3 and 5). It is also observed that the detected signal is more sensitive to phase variations than to power variations. Thus, cases 6 and 7 show that a relative amplitude variation of 5:1 (0.1/0.02) with respect to a relative phase variation of 1.32:1 (0.99/0.75), provides the same power drop (-41.19 dB).

\begin{table}[!t]
\caption{Some values that relate the value of the null measured in the SA (Spectrum Analyzer), to the amplitude and phase deviations of the secondary generator signal.}
\label{tab:table_i}
\centering
\begin{tabular}{|c|c|c|c|c|c|c|}
\hline
\multirow{2}{*}{\textrm{Case}} & $\mathit{a}$ & $\mathit{s}$ & $\mathit{\Delta\theta_S}$ &
$\mathit{SA_{max}}$ & $\mathit{SA_{min}}$ & $\mathit{r}$ \\
 & \textrm{(dBm)} & \textrm{(dB)} & \textrm{(\textordmasculine)} & \textrm{(dBm)} &
\textrm{(dBm)} & \textrm{(dB)} \\ \hline
1 & 0 & 0 & 0 & 6.02 & -$\infty$ & -$\infty$ \\ \hline
2 & 0 & 0.15 & 0 & 6.10 & -35.18 & -41.18 \\ \hline
3 & 0 & 0 & 1 & 6.02 & -35.16 & -41.18 \\ \hline
4 & 5 & 0.15 & 0 & 11.10 & -35.18 & -41.28 \\ \hline
5 & 5 & 0 & 1 & 11.02 & -30.16 & -41.18 \\ \hline
6 & 0 & 0.02 & 0.99 & 6.03 & -35.16 & -41.19 \\ \hline
7 & 0 & 0.1 & 0.75 & 6.07 & -35.12 & -41.19 \\
\hline
\end{tabular}
\end{table}

In practice, when the E8257D and E8257C generators share the 10 MHz reference signal, there is a very small frequency difference between them ($\Delta w$ $\approx$ 0 in Fig. \ref{fig:figura1}), which results in a slow variation of the phase shift over time. In that case, the generators are considered to be quasi-synchronized. That frequency variation is greater the higher the frequency selected in the generators. With the null technique it is observed that the output signal at the power combiner successively increases until it reaches the maximum and decreases until it reaches the minimum. Despite this continuous phase variation, it is possible to perform the characterization of a phase detector using the maximum permissible error specification. However, this phase variation means that part of the calibration and measurement procedure must be carried out jointly, as will be explained in Section VI.

\section{The calibration and measurement system}
\begin{figure}[!t]
\centering
\includegraphics[width=7.9cm]{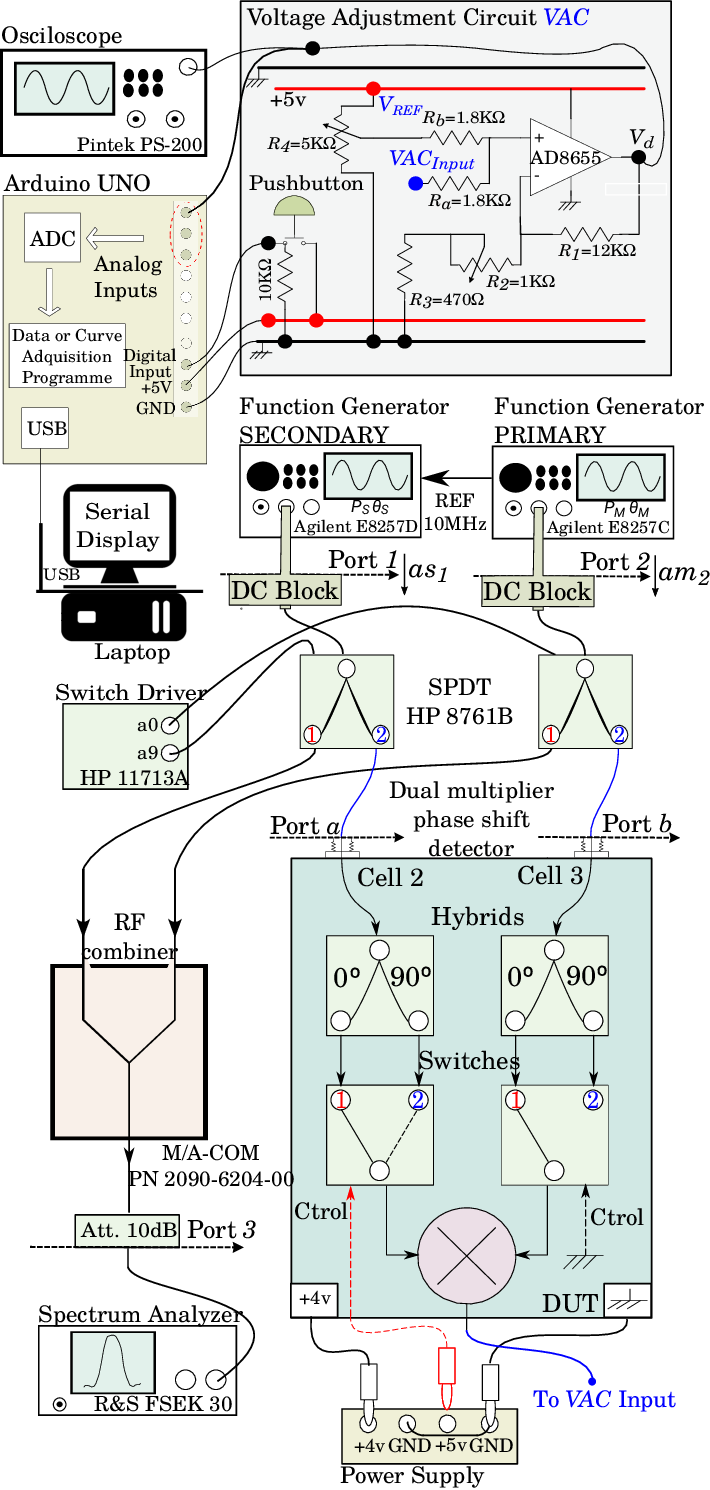}
\caption{Simplified diagram of the calibration and measurement system.}
\label{fig:figura3}
\end{figure}

Fig. \ref{fig:figura3} shows the calibration and measurement system block diagram. It is mainly composed of two generators (E8257D and E8257C) that share the 10 MHz reference signal, a power combiner, the phase detector or DUT, two switches that direct the signal from the generators to the inputs of the combiner, an SA (R\&S FSEK 30) to measure the output power of the combiner, an Arduino Uno to semi-automate the procedure, a $\mathit{VAC}$ (Voltage Adjustment Circuit) to adapt the output signals of the phase detector to the dynamic range of the Arduino Uno ADC, and an oscilloscope to observe the phase detector response. The generators work as primary (E8257C) and secondary (E8257D). They are characterized by their power ($P_M$ and $P_S$) and phase ($\theta_M$ and $\theta_S$). The switches (HP 8761B) are controlled by two buttons (a0 and a9), integrated into the HP 11713A driver.

The input voltage of Arduino Uno ADC must be between 0 and 5 V. Therefore, when the output signal of the phase detector is bipolar, a positive voltage ($V_{REF}$) must be added to the output voltage of the phase detector. In addition, an amplifier network ($R_1$, $R_3$ and the potentiometer $R_2$) has been introduced to maximize the input levels to the ADC.  As a result, the output voltage of the $\mathit{VAC}$ is given by (Fig. \ref{fig:figura3}):

\begin{equation}
V_d = \frac{V_{ref} + \mathit{VAC}_{Input}}{2} \cdot \left[1+\frac{R1}{R2+R3}\right]
\end{equation}

\section{The DUT: 360\textordmasculine\ phase detector based on switched dual multiplier}
A switched dual multiplier phase detector has been selected to explain the calibration and measurement procedure. The characterization of this type of detectors consists of obtaining two curves: in-phase and phase-shifted \cite{Perez2021, Perez2017}. For simplicity reasons, they will be named $IQ$, although this type of detector is not required to strictly fulfill the quadrature condition, that is, the phase difference of 90\textordmasculine\ between the input signals for product $Q$.

\begin{figure}[!t]
\centering
\subfloat[]{\includegraphics[width=4.3cm]{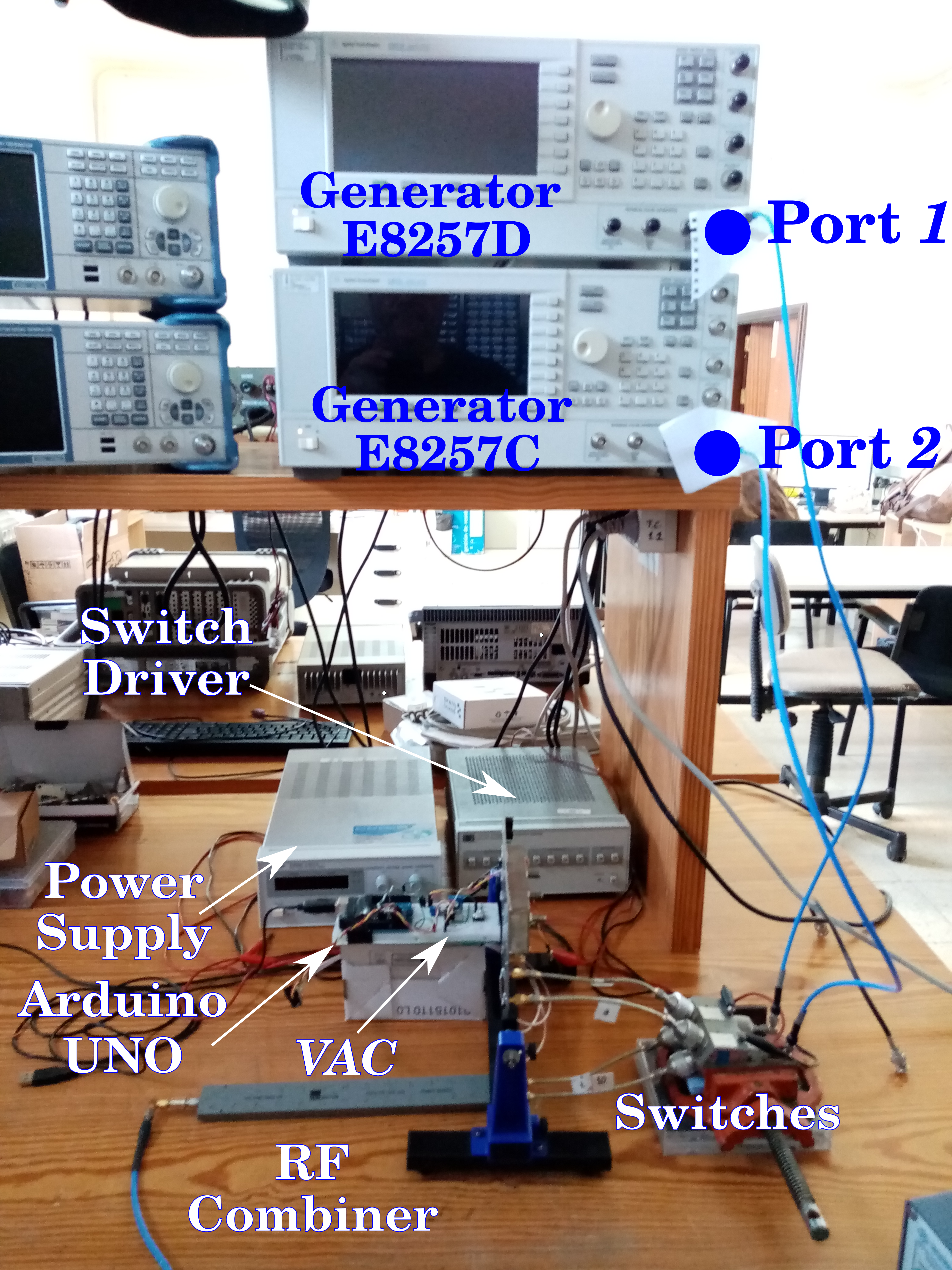}%
\label{fig:figura4a}}
\hspace{0.0cm}
\subfloat[]{\includegraphics[width=4.3cm]{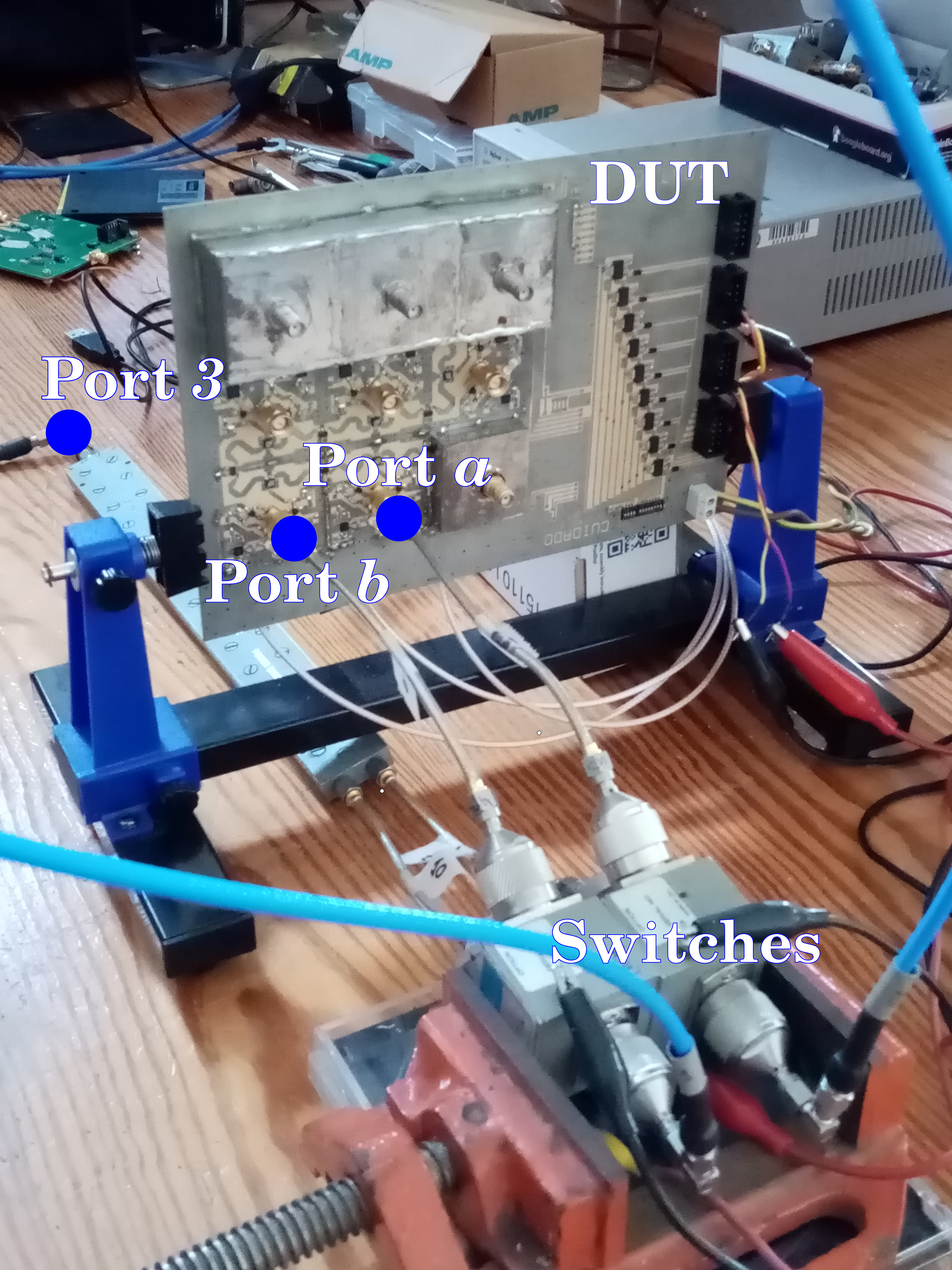}%
\label{fig:figura4b}}
\caption{\textbf{(a)} Photograph showing devices and equipment used for calibration and measurement. \textbf{(b)} Photograph showing the connection from the switches to the power combiner and the phase detector (cells 2 and 3 of the measuring array). The SA and the oscilloscope are out of frame.}
\label{fig:figura4}
\end{figure}

The simplified diagram is shown in Fig. \ref{fig:figura3} (dual multiplier phase shift detector block), and Fig. \ref{fig:figura4} shows a photograph of the measurement system, as well as a photograph of the DUT and power combiner connections. It is composed of two identical cells containing a hybrid circuit and a switch. The selected signals enter a multiplier, although the switch in cell 3 is always at low level, position 1). In appearance, a $I\times$$I$ multiplication is performed when the switch in cell 2 is in position 1), and a $Q\times$$I$ multiplication when it is in position 2). However, the operation described in \cite{Perez2017} demonstrates that it can be used as a 360\textordmasculine\ phase detector, despite there being a maximum theoretical deviation of $\pm$90\textordmasculine\ over the quadrature condition. In practice, that deviation is linked to the maximum admissible error in phase detection and, therefore, has values lower than the theoretical value. In other words, the hybrid circuit can operate in the range 90$\pm\beta$\textordmasculine, with $\beta$ being a value less than 90\textordmasculine.

The objective of this paper is to show the measurement procedure through quasi synchronized generators, so an arbitrary value $\beta$ will be fixed as a condition to know if the output curves of the detector comply with that specification. Moreover, these output voltage curves refer to a power and frequency that must be fixed in advance.

Regarding the sign convention when calculating the phase difference between the output voltages of the phase detector, the following considerations shall have to be taken into account. The hybrid circuit (QCS-592) used in the double multiplication phase detector introduces a phase shift of +90\textordmasculine\ compared to the direct port. The sign convention is maintained for the primary and secondary generators of Section II. As a first approximation, if paths are balanced, the low frequency component of the $I\times$$I$ product is given by (\ref{eq:wt}):
\begin{equation}
\label{eq:wt}
\begin{split}
& \cos(wt) \cdot \cos(wt+\theta_M) \\
&=(1/2)\cdot\left[\cos(2wt+\theta_M)+\cos(wt-wt-\theta_M)\right] \\
&\simeq(1/2)\cdot\cos(-\theta_M)=(1/2)\cdot\cos(\theta_M)
\end{split}
\end{equation}
For the $Q\times$$I$ path:
\begin{equation}
\label{eq:wt90}
\small
\begin{split}
& \cos(wt+90)\cdot \cos(wt+\theta_M) \\
&=(1/2)\cdot\left[\cos(2wt+90+\theta_M) +\cos(wt+90-wt-\theta_M)\right] \\
&\simeq(1/2)\cdot\cos(\theta_M-90)
\end{split}
\end{equation}

\begin{figure}[!t]
\centering
\subfloat[]{\includegraphics[width=7.4cm]{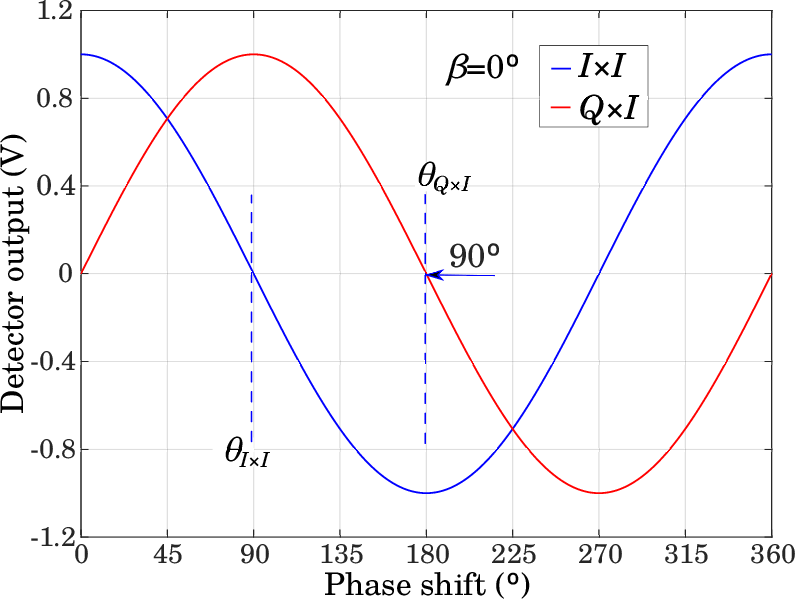}%
\label{fig:figura5a}}
\hspace{0.0cm}
\subfloat[]{\includegraphics[width=7.4cm]{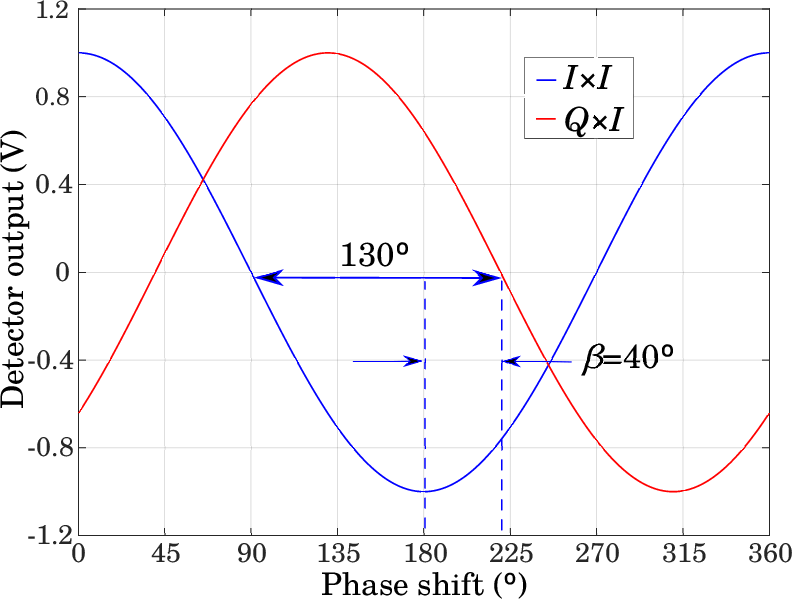}%
\label{fig:figura5b}}
\caption{$I\times$$I$ and $Q\times$$I$ output signals of the detector to establish the sign convention, $\theta_{Q\times I}$ - $\theta_{I\times I}$. \textbf{(a)} 90\textordmasculine\ phase shift ($\beta$ = 0\textordmasculine) and \textbf{(b)} 130\textordmasculine\ phase shift ($\beta$ $\approx$ 40\textordmasculine).}
\label{fig:figura5}
\end{figure}

Fig. \ref{fig:figura5} represents the $I\times$$I$ and $Q\times$$I$ signals. It shows the sign convention that will be used to calculate the phase shift between the two curves of the detected voltage depending on the phase shift of the input signals. The convention $\theta_{Q\times I}$ - $\theta_{I\times I}$ will be used, being +90 when in quadrature. Therefore, if it were acceptable to have a deviation of $\beta =$ 40\textordmasculine\ regarding the quadrature condition, the phase difference between curves must be between +90-40\textordmasculine\ and +90+40\textordmasculine. Fig. \ref{fig:figura5b} depicts a case when the phase shift is 130\textordmasculine\ (90+40\textordmasculine).

\section{Errors due to unbalanced paths: VNA calibration}
If the set of devices connecting the generators to the DUT and the SA is not considered to be well balanced enough, it is advisable to perform a calibration and correct the parameters of each generator accordingly. This ensures that the input signals to the DUT keep the defined phase shift between generators and that they have the same amplitude.

In this case, and to highlight the correction procedure, this type of calibration is carried out. The $S$ parameters of the connection block with 5 ports are measured: $\mathit{1}$, $\mathit{2}$, $\mathit{3}$, $a$ and $b$ (Fig. \ref{fig:figura3}). Under conditions of matching and good isolation between ports, it is enough to measure the following parameters: $S_{31}$ and $S_{32}$ with 8761B switches in state 2);  $S_{a1}$ and $S_{b2}$ with 8761B switches in state 1).

When the generators have been adjusted to obtain a null at SA, the signals at the output of the combiner are in counterphase ($\Delta\theta =$ 180\textordmasculine) and possess equal amplitude. If $as_1$ and $am_2$ are the power waves of the generators and are a good output matching (10 dB attenuator at the output of the power combiner), then the following is met:
\begin{equation}
\label{eq:as1}
as_1 \cdot \frac{S_{31}}{am_2} \cdot S_{32} = |1|_{180^o}\Longrightarrow am_2 = as_1 \cdot \frac{S_{31}}{S_{32} \cdot |1|_{180^o}}
\end{equation}
If the phase detector inputs are considered to be matched, the null condition must still be maintained when switching the signals from the generators to the phase detector, even though this requires a slight modification of the amplitude and phase of the generators, i.e.
\begin{equation}
\label{eq:as1prime}
as_1^\prime \cdot \frac{S_{a1}}{am_2^\prime}\cdot S_{b2} = |1|_{180^o} \Longrightarrow   as_1^\prime = am_2^\prime \cdot |1|_{180^o} \cdot \frac{S_{b2}}{S_{a1}}
\end{equation}
If the parameters of the primary generator are not modified ($am_2^\prime$ = $am_2$), from (\ref{eq:as1}) and (\ref{eq:as1prime}) yields:
\begin{equation}
\label{eq:as1primefinal}
as_1^\prime = as_1 \cdot \frac{S_{31} \cdot S_{b2}}{S_{32} \cdot S_{a1}} = as_1 \cdot |\Delta {A_S}_c|_{\Delta{\theta_S}_c}
\end{equation}

Therefore, when the signals to the phase detector are switched, the amplitude and phase of the secondary generator must be corrected to $\Delta {A_S}_c$ ($\Delta {P_S}_c$ dB) and $\Delta {\theta_S}_c$, respectively. To make it easier for manual change, it is better to modify the secondary generator to $\Delta{\theta_S}_c$ and the primary generator to $\Delta {P_M}_c = - \Delta {P_S}_c$ dB. Table \ref{tab:table_ii} shows the correction values for the frequencies tested, as well as the $S$ parameters of the connection block. $S$ parameters has been measured with the R\&S ZVK VNA.

\begin{table}[!t]
\caption{Relevant values in the measurement process: $S$ parameters of the connection block (Fig. \ref{fig:figura3}) and correction values related to the power and phase of generators.}
\label{tab:table_ii}
\centering
\begin{tabular}{|c|c|c|c|c|c|c|}
\hline
$\mathit{f}$ & $\mathit{S_{31}}$ & $\mathit{S_{32}}$ & $\mathit{S_{a1}}$ &
$\mathit{S_{b2}}$ & $\mathit{\Delta{P_M}_c}$ & $\mathit{\Delta{\theta_S}_c}$\\
\textrm{(GHz)} & \textrm{(dB)/(\textordmasculine)} & \textrm{(dB)/(\textordmasculine)} &
\textrm{(dB)/(\textordmasculine)} & \textrm{(dB)/(\textordmasculine)} &
\textrm{(dB)} & \textrm{(\textordmasculine)}\\ \hline
\multirow{2}{*}{3} & -10.71 & -10.64 & -1.28 & -1.27 & \multirow{2}{*}{-0.066} & \multirow{2}{*}{-0.331}\\
  & 119.71 & 121.27 & 77.68 & 78.90 & & \\ \hline
\multirow{2}{*}{4} & -11.24 & -11.12 & -1.71 & -1.57 & \multirow{2}{*}{0.017} & \multirow{2}{*}{-0.331}\\
  & 43.27 & 45.04 & -134.35 & -133.53 & & \\ \hline
\multirow{2}{*}{5} & -11.56 & -11.57 & -1.81 & -1.86 & \multirow{2}{*}{-0.047} & \multirow{2}{*}{-0.437}\\
  & -33.54 & -32.94 & 14.51 & 14.68 & & \\ \hline
\multirow{2}{*}{6} & -11.73 & -11.71 & -1.85 & -2.00 & \multirow{2}{*}{-0.172} & \multirow{2}{*}{-0.470}\\
  & -110.95 & -109.39 & 162.22 & 163.31 & & \\ \hline
\multirow{2}{*}{7} & -12.09 & -12.05 & -1.98 & -2.12 & \multirow{2}{*}{-0.178} & \multirow{2}{*}{-0.164}\\
  & 171.06 & 173.02 & -50.95 & -49.15 & & \\ \hline
\multirow{2}{*}{8} & -12.49 & -12.34 & -2.18 & -2.31 & \multirow{2}{*}{-0.281} & \multirow{2}{*}{-0.544}\\
  & 93.83 & 95.32 & 97.05 & 99.09 & & \\
\hline
\end{tabular}
\end{table}

If the calibration due to path unbalance were not duly conducted, and considering the phase correction data ($\Delta{\theta_S}_c$), in the worst-case scenario, the uncertainty of an error would increase by almost one degree more (-0.94\textordmasculine\ at 4 GHz in Table \ref{tab:table_ii}) than the one defined in the generator calibration stage (Table \ref{tab:table_i}). The 0.3 dB ($f=$ 8 GHz) variation in power has little influence on the response of the phase detector, so it is not that relevant.

However, -0.94\textordmasculine\ is within the transmission accuracy specifications of the R\&S ZVK VNA provided by the manufacturer \cite{1EZ48}: $\pm$1\textordmasculine\ and $\pm$0.1 dB between 2 GHz and 8 GHz. In other words, this data shows that the use of two generators facilitates the symmetry of the signal distribution network, which reduces imbalance and makes the use of a VNA not mandatory. Although, if the measurement error is to be bounded more rigorously, the VNA should always be used to guarantee the specific value derived from equation (\ref{eq:as1primefinal}). Despite all the above, the correction values have been included in the procedure to show how they should be taken into account.

\section{Calibration and measurement procedure}
The calibration and measurement procedure is divided into two clearly differentiated parts: data collection to reference the $I\times$$I$ and $Q\times$$I$ curves (reference data), and data collection of the data from the $I\times$$I$ and $Q\times$$I$ curves (curve data). In addition, the process for acquiring the reference data contains a block of tasks that should be executed in the shortest possible time. Generally speaking, the idea is to collect the reference data while the phase shift of the generator is kept within a specified error (Table \ref{tab:table_i}). Each part has an associated program that activates the data collection through a pushbutton on the breadboard, next to the $\mathit{VAC}$ (Fig. \ref{fig:figura3}).

\begin{figure}[!t]
\centering
\includegraphics[width=7.5cm]{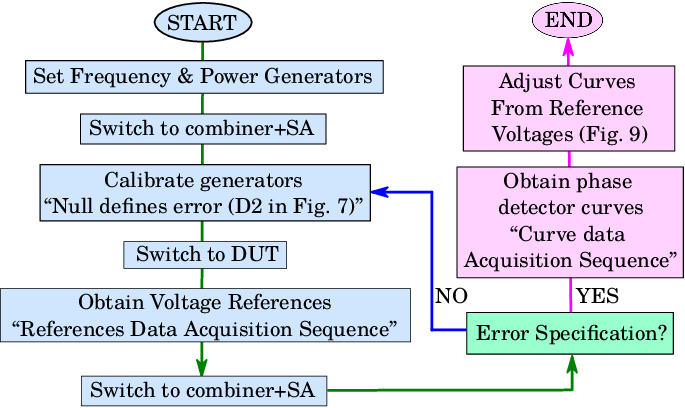}
\caption{Simplified flow chart that illustrates the measurement and calibration process.}
\label{fig:figura6}
\end{figure}

The output amplitude of the multiplier is adjusted at each frequency so that the ADC works at around 0-5 V. It is necessary to ensure that the $I\times$$I$ and $Q\times$$I$ curves do not exceed these values. Table \ref{tab:table_iii} includes the correction values of the generators (one decimal versus three in Table \ref{tab:table_ii}) and the most relevant data from the calibration and measurement process (Fig. \ref{fig:figura3}): ${V_d}_{max}$ (maximum output voltage of the detector for $I\times$$I$), ${V_d}_{min}$ (minimum voltage), $\Delta {P_M}_{null}$ (power increase in the primary generator over that of the secondary to attain a null at the output of the power combiner), $P_{SUM}$ (maximum power at the combiner output), $\Delta {P_M}_c$ (power correction at the primary generator), $\Delta{\theta_S}_c$ (phase correction at the secondary generator), and $\Delta P$ (power correction to reference in Port \textit{a} and \textit{b}). The measurement and calibration process is summarized in Fig. \ref{fig:figura6} and detailed below for a power of 0 dBm in the generators and an error of $\pm$1\textordmasculine:\\

\begin{figure}[!t]
\centering
\includegraphics[width=7.2cm]{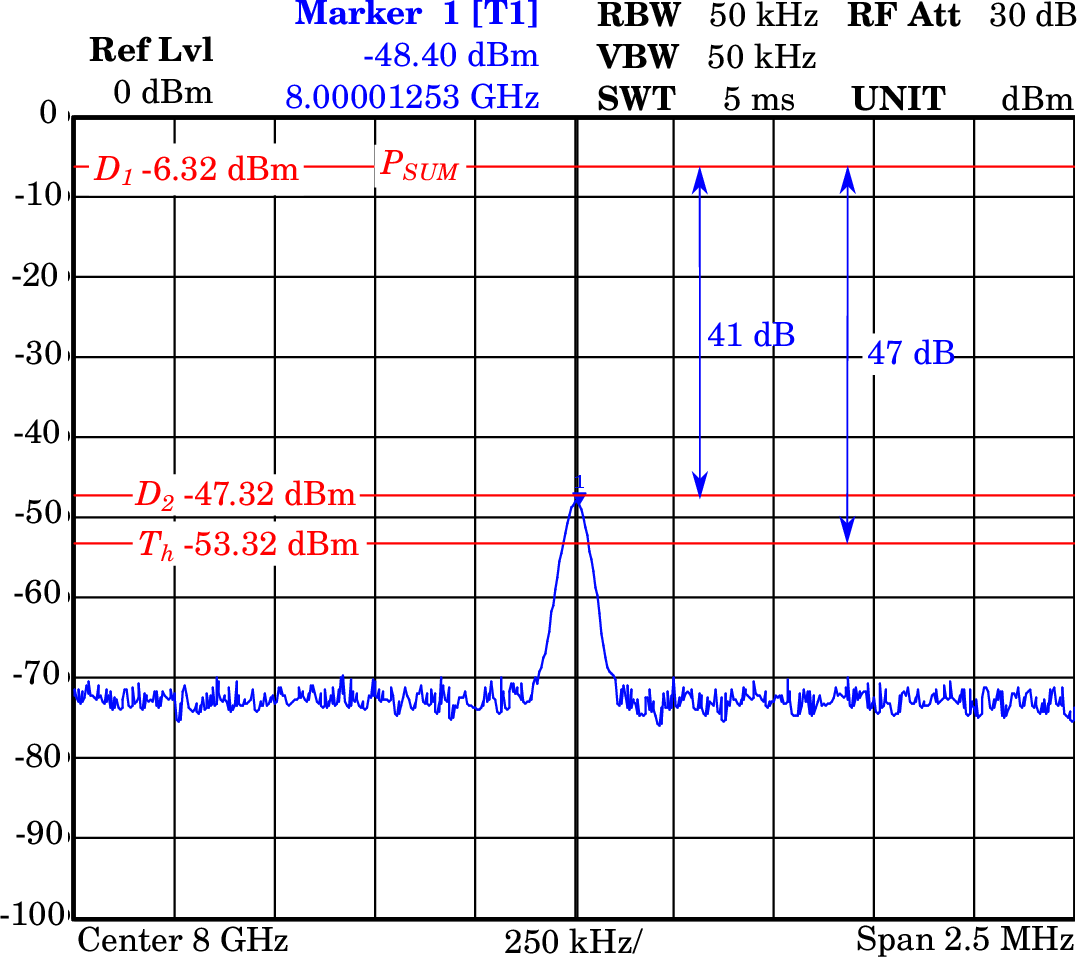}
\caption{SA image when the signal at the output of the combiner is just below the threshold at -41dB of the $P_{SUM}$ value. It shows the reference lines to $P_{SUM}$, $P_{SUM}$ - 41dB ($\epsilon = \pm$1\textordmasculine) and $P_{SUM}$ - 47dB ($\epsilon = \pm$0.5\textordmasculine) at a frequency of 8 GHz.}
\label{fig:figura7}
\end{figure}

{\renewcommand\labelitemi{$\cdot$}
\begin{itemize}
\small
\item[]\textit{Main sequence of combined calibration/measurement}
 \item \textbf{Set} frequency and power generators: $P_M = P_S =$ 0 dBm.
 \item \textbf{Connect} red banana connector to GND ($I\times$$I$).
 \item \textbf{Push} a0 and a9 to switch SPDT 8761B to 1), towards DUT.
 \item \textbf{Increase} primary generator frequency by 1 kHz.
 \item \textbf{Visualize} the detector output voltage ($V_d$) in oscilloscope.
 \item \textbf{Adjust} $R_4$ and $R_2$ potentiometers to maximize $V_d$.
 \item \textbf{Decrease} primary generator frequency by 1 kHz.
 \item \textbf{Push} a0 and a9 to switch SPDT 8761B to 2), towards RF combiner.
 \item [\textcolor{red}{\ding{172}}] \textbf{Set} $\theta_M =$ 180\textordmasculine\ and adjust $\theta_S$ (${\theta_S}_{REF}$) and $P_M$ ($\Delta {P_M}_{null}$) to get minimum power in SA.
 \item \textbf{Set} $\theta_M =$ 0\textordmasculine\ and measure maximum power ($P_{SUM}$) in SA.
 \item \textbf{Set} a line reference at $P_{SUM}$ - 41 dB in SA.
 \item \textbf{Set} $\theta_M =$ 180\textordmasculine\ and adjust $\theta_S$ to get SA power close to line reference.
 \item \textbf{Ensure} that the phase evolves so that the SA power decreases with time.
 \item \textbf{Start} \textit{Reference Data Acquisition} (Detailed below) since the SA power is clearly under line reference (Fig. \ref{fig:figura7}).
 \item \textbf{Check} SA power when Data Acquisition sequence is finished.
 \item \textbf{If} SA power continues under line reference (Fig. \ref{fig:figura7}), Data is right.
 \item \textbf{Otherwise}, it is necessary repeat this sequence from \textcolor{red}{\ding{172}}.
\end{itemize}
}

\begin{table}[!t]
\caption{Relevant values in the measurement process. For all cases, the power of the secondary generator is  $P_S =$ 0 dBm.}
\label{tab:table_iii}
\resizebox{0.5\textwidth}{!}
{
\begin{tabular}{|c|c|c|c|c|c|c|c|}
\hline
$\mathit{f}$ &
\makebox[0.65cm]{$\mathit{{V_d}_{max}}$} &
\makebox[0.65cm]{$\mathit{{V_d}_{min}}$} &
\makebox[0.9cm]{$\mathit{\Delta{P_M}_{null}}$} &
\makebox[0.6cm]{$\mathit{P_{SUM}}$} &
\makebox[0.6cm]{$\mathit{\Delta{P_M}_c}$} &
\makebox[0.6cm]{$\mathit{\Delta{\theta_S}_c}$} &
\makebox[0.6cm]{$\mathit{\Delta P}$}\\
\makebox[0.6cm]{\textrm{(GHz)}} & \textrm{(V)} & \textrm{(V)} & \textrm{(dB)} &
\textrm{(dBm)} & \textrm{(dB)} & \textrm{(\textordmasculine)} &
\textrm{(dB)} \\ \hline
3 & 4.77 & 0.46 & -0.2 & -9.5  & 0.1 & -0.3 & -1.3 \\ \hline
4 & 4.46 & 0.26 &  0   & -9.47 & 0   & -0.9 & -1.7 \\ \hline
5 & 4.91 & 0.13 &  0   & -8.03 & 0   & -0.4 & -1.8 \\ \hline
6 & 4.82 & 0.16 & -0.1 & -7.78 & 0.2 & -0.5 & -1.8 \\ \hline
7 & 4.76 & 0.11 & -0.3 & -7.43 & 0.2 & -0.2 & -2.0 \\ \hline
8 & 4.45 & 0.48 & -0.1 & -6.32 & 0.3 & 0.5  & -2.2 \\
\hline
\end{tabular}
}
\end{table}

The block of tasks, named \textit{Reference Data Acquisition}, needs to be performed as quickly as possible to ensure that the measurements are correct. The higher the frequency, the shorter the time available. However, if the errors derived from Table \ref{tab:table_ii} are accepted, the step to adjust the values of $P_M$ and $\theta_S$ can be skipped. In that case, a good estimate for the error is to increase its value by $\left|\Delta{\theta_S}_c\right|$, in other words, $\epsilon = \pm\left(1+\left|\Delta{\theta_S}_c\right|\right)$, if the reference line is at 41 dB. The sequence of the measurement procedure in order to obtain the reference voltages ($V_d$ at 180\textordmasculine\ and 90\textordmasculine\ when $Q\times$$I$ or $I\times$$I$ is selected) is as follows:

{\renewcommand\labelitemi{$\cdot$}
\begin{itemize}
\small
 \item[]\textit{Reference Data Acquisition sequence}
 \item \textbf{Push} a0 and a9 to switch SPDT 8761B to 1), towards DUT.
 \item \textbf{Adjust} $P_M$ with $\Delta{P_M}_c$ and ${\theta_S}_{REF}$ with $\Delta{\theta_S}_c$ (Table \ref{tab:table_iii}).
 \item \textbf{Press} the pushbutton to acquire $V_d$ at 180\textordmasculine\ and $I\times$$I$ ($\mathit{Vi}_{180}$).
 \item \textbf{Change} $\theta_M$ to 90\textordmasculine\ and press the pushbutton ($\mathit{Vi}_{90}$).
 \item \textbf{Connect} red banana connector to +5 V ($Q\times$$I$) ($\mathit{Vq}_{90}$).
 \item \textbf{Press} the pushbutton to acquire $V_d$ at 90\textordmasculine\ and $Q\times$$I$ ($\mathit{Vq}_{180}$).
 \item \textbf{Change} $\theta_M$ to 180\textordmasculine\ and press the pushbutton.
 \item \textbf{Restore} $P_M$ with $\Delta{P_M}_c$ and ${\theta_S}_{REF}$ with $\Delta{\theta_S}_c$ (Table \ref{tab:table_iii}).
 \item \textbf{Push} a0 and a9 to switch SPDT 8761B to 2) towards RF combiner.
\end{itemize}
}

After that, the program for the acquisition of the curve data ($I\times$$I$ and $Q\times$$I$) is started, and the following set of tasks is performed:

{\renewcommand\labelitemi{$\cdot$}
\begin{itemize}
\small
 \item[]\textit{Curve data acquisition sequence}
 \item \textbf{Adjust} $P_M$ with $\Delta{P_M}_c$.
 \item \textbf{Increase} primary generator frequency by 11 Hz.
 \item \textbf{Connect} red banana connector to GND ($I\times$$I$).
 \item \textbf{Press} the pushbutton to acquire $I\times$$I$ curve.
 \item \textbf{Connect} red banana connector to +5 V ($Q\times$$I$).
 \item \textbf{Press} the pushbutton to acquire $I\times$$I$ curve.
 \item \textbf{Decrease} primary generator frequency by 11 Hz.
\end{itemize}
}

\begin{figure}[!t]
\centering
\includegraphics[width=7.2cm]{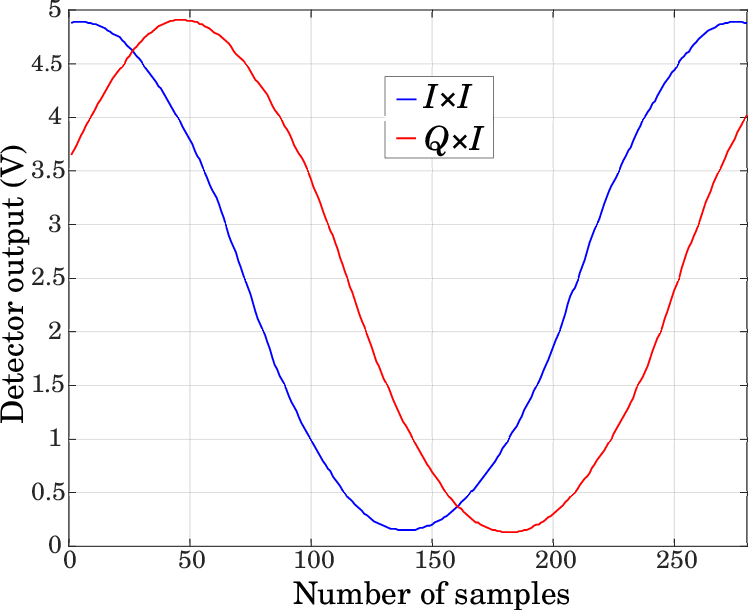}
\caption{Curves representing the output voltages of the phase detector when $Q\times$$I$ (blue) and $I\times$$I$ (red) are multiplied. The difference in frequency between the generators is 11 Hz, and each curve is composed of 280 samples.}
\label{fig:figura8}
\end{figure}

The 11 Hz frequency difference between generators ensures that 280 samples are enough to have more than one period and to not exceed both ADC converter speed and the available memory on Arduino Uno (Fig. \ref{fig:figura8}). At this point, the 4 reference voltages ($V_d$ at 180\textordmasculine\ and 90\textordmasculine\ when $Q\times$$I$ and $I\times$$I$ is selected) and slightly more than one period of the $I\times$$I$ and $Q\times$$I$ curves are available.

\section{Curve referencing}
\begin{figure}[!t]
\centering
\subfloat[]{\includegraphics[width=7.5cm]{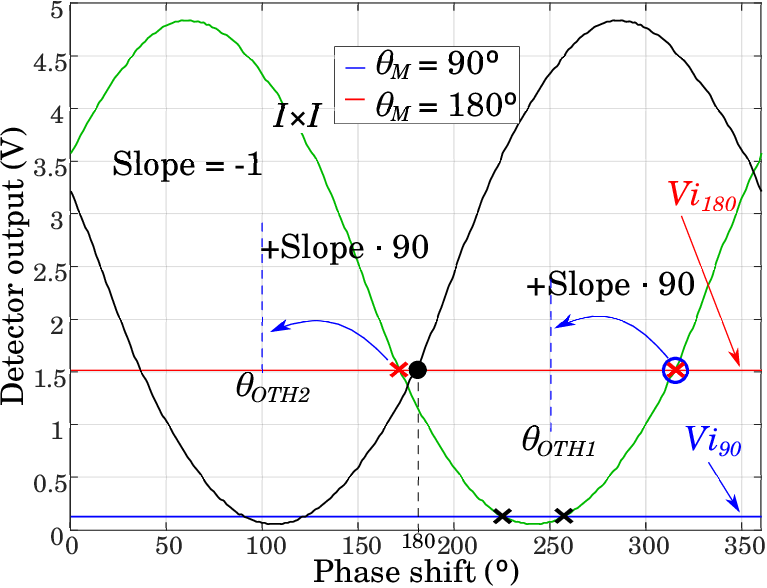}%
\label{fig:figura10a}}
\hspace{0.0cm}
\subfloat[]{\includegraphics[width=7.5cm]{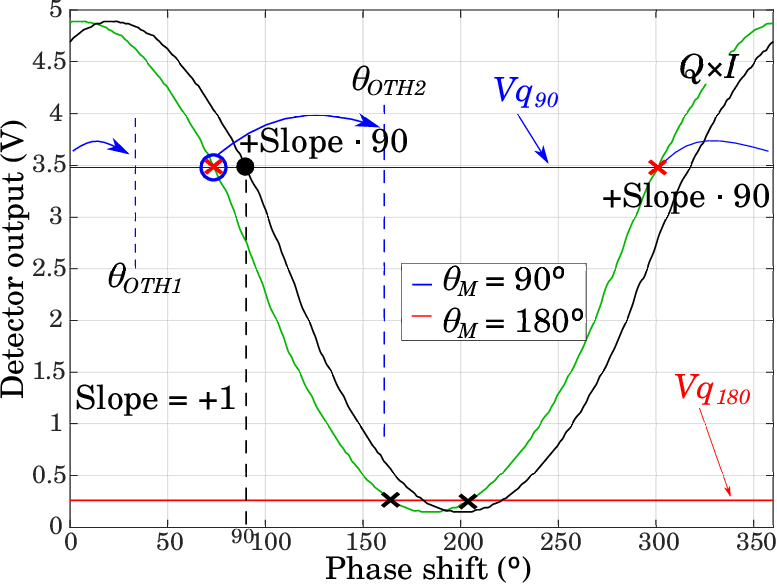}%
\label{fig:figura10b}}
\caption{\textbf{(a)} Representation of the curves and data of interest to reference the curves of the phase detector output voltages. It shows: a period obtained from the original curve ($I\times$$I$ in Fig. \ref{fig:figura8}), in green; the phase detector voltage when phase shift $\theta_M =$ 90\textordmasculine\ (blue) and $\theta_M =$ 180\textordmasculine\ (red); the referenced curve (black) restored from the solution ${\theta_{REF}}_{180} =$ 315.000\textordmasculine\ (blue ``O'' on the green curve). \textbf{(b)} Similar curves and values to reference the $Q\times$$I$ curve in Fig. \ref{fig:figura8}, where ${\theta_{REF}}_{90} =$ 73.667\textordmasculine.}
\label{fig:figura10}
\end{figure}

The data sent by the serial port to the computer is processed to obtain correctly referenced $I\times$$I$ and $Q\times$$I$ curves. These curves characterize the double multiplier phase detector and allow the calculation of the phase difference between them and whether they exceed a certain target value (90$\pm\beta_{max}$\textordmasculine) \cite{Perez2021, Perez2017}. The procedure for referencing curves takes into account the sign conventions established in Sections II and IV.
Fig. \ref{fig:figura9} details the referencing process whose main curves and related data are shown in Fig. \ref{fig:figura10}.

\begin{figure}[!t]
\centering
\includegraphics[width=7.6cm]{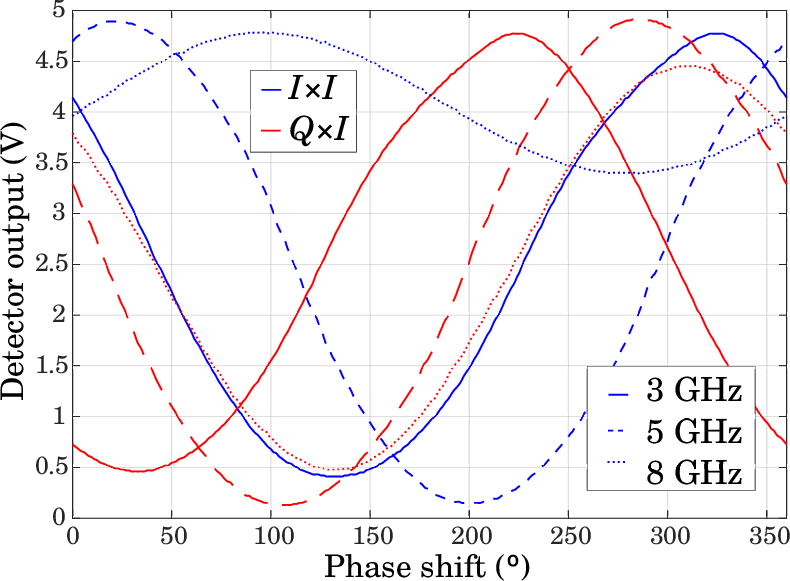}
\caption{$I\times$$I$ (blue) and $Q\times$$I$ (red) curves for three frequency values: 3 GHz, 5 GHz and 8 GHz.}
\label{fig:figura11}
\end{figure}

Fig. \ref{fig:figura11} shows the $I\times$$I$ and $Q\times$$I$ curves referenced for 3 GHz, 5 GHz and 8 GHz. A large amplitude difference is observed between the curves at 8 GHz, primarily due to being significantly outside the band of the hybrid circuit (QCS-592: 3.1 GHz - 5.9 GHz) which is not a problem for this type of detector \cite{Perez2021}. Table \ref{tab:table_iv} provides the reference values for each frequency and curve ($I\times$$I$ and $Q\times$$I$): \textit{y} (value of $\theta_M$ that provides an output voltage of the phase detector closer to the mean value), ${\theta_{REF}}_y$ (reference value to be used for reconstructing the output voltage curve of the phase detector), ${V_{REF}}_y$ (output voltage of the detector when $\theta_M = y$), and curve ($Q\times$$I$ or $I\times$$I$).  Lastly, the offset between the curves $Q\times$$I$ and $I\times$$I$ is obtained and it is determined if it is within the range bounded by the maximum deviation, that is, 90$\pm\beta_{max}$\textordmasculine. The last two columns of Table \ref{tab:table_iv} show the value of the phase difference between the curves, specifying whether they are within the specified range.

\begin{figure}[!t]
\begin{algorithmic}[1]
\small
\STATE \textbf{Initialize} (curve$_i$, curve$_q$, \COMMENT{$i$ refers to $I\times$$I$ and $q$ refers to $Q\times$$I$}\\
\hspace{1.2cm} $\mathit{Vi}_{180}$, $\mathit{Vi}_{90}$, $\mathit{Vq}_{90}$ and $\mathit{Vq}_{180}$) \COMMENT{reference voltages}.
\STATE \textbf{Oversample} (curves by 4 to minimize errors)
\STATE \textbf{Calculate} (period and phase vector) \COMMENT{green in Fig. \ref{fig:figura10}}
\STATE \textbf{For} each curve ($x$ = $i$, $q$)\\
\STATE \hspace{0.2cm} \textbf{Select} $V_{REF_y}$ closed to curve middle value ($V_{dm}$).
\STATE \hspace{0.4cm}  \textbf{If} $\mathit{Vx}_{180}$\\
\STATE \hspace{0.6cm} ${V_{REF}}_y = \mathit{Vx}_{180}$, Slope $=$ -1, $V_o = \mathit{Vx}_{90}$\\
\STATE \hspace{0.4cm}  \textbf{If} $\mathit{Vx}_{90}$\\
\STATE \hspace{0.6cm} ${V_{REF}}_y = \mathit{Vx}_{90}$, Slope $=$ +1, $V_o = \mathit{Vx}_{180}$\\
\STATE \hspace{0.2cm} \textbf{Calculate} phases where ${V_{REF}}_y$ cuts curve ($y$ $=$ 90 or 180).\\
\hspace{0.8cm} \COMMENT {${\theta_{REF}}_{y1}$ and ${\theta_{REF}}_{y2}$ (red ``$\times$'' in Fig. \ref{fig:figura10})}\\
\hspace{0.2cm} \COMMENT{Solve the ambiguity to obtain the solution.}\\
\STATE \hspace{0.5cm} \textbf{Estimate} the other $\theta_{REF}$ from ${\theta_{REF}}_{yk}$  ($k=$ 1 and 2).\\
\hspace{1.1cm} \COMMENT {${\theta_{OTHyk}} = {\theta_{REF}}_{yk}$ + Slope $\cdot$ 90}.\\
\STATE \hspace{0.5cm} \textbf{Calculate} phases where $V_o$ intersects curve.\\
\hspace{1.1cm} \COMMENT {${\theta_O}_1$ and ${\theta_O}_2$ (black ``$\times$'' in Fig. \ref{fig:figura10})}.\\
\STATE \hspace{0.5cm} \textbf{Compare} ${\theta _{OTHyk}}$ with ${\theta_O}_k$.\\
\STATE \hspace{0.5cm} \textbf{Calculate} solution from the closest difference to zero. \\
\hspace{1.1cm} \COMMENT {${\theta_{REF}}_{yk}$ ($k =$ 1 or 2) (blue ``O'' in Fig. \ref{fig:figura10})}\\
\STATE \hspace{0.2cm} \textbf{Restore} the curve from from  previous solution.\\
\hspace{0.8cm} \COMMENT {black curve from black dot in Fig.\ref{fig:figura10}}.
\end{algorithmic}
\caption{Simplified algorithm to reference the $I\times$$I$ and $Q\times$$I$ curves.}
\label{fig:figura9}
\end{figure}

\begin{table}[!t]
\caption{Values obtained to reference the curves at evaluated frequencies, phase shift between $I\times$$I$ and $Q\times$$I$ curves and fulfillment of 90$\pm$40\textordmasculine\ bounding.}
\label{tab:table_iv}
\centering
\begin{tabular}{|c|c|c|c|c|c|c|}
\hline
$\mathit{f}$ & $\mathit{y}$ & $\mathit{\theta_{REF_y}}$ &
$\mathit{V_{REF_y}}$ & \textrm{Curve} &
\textrm{Phase shift} & \textrm{90$\pm$40}\\
{\textrm{(GHz)}} & \textrm{(\textordmasculine)} & \textrm{(\textordmasculine)} & \textrm{(V)} & \textrm{Type} & \textrm{(\textordmasculine)} & \textrm{(\textordmasculine)}\\ \hline
\multirow{2}{*}{3} & 180 & 192.000 & 0.950 & $Q\times$$I$ &
\multirow{2}{*}{99.967} & \multirow{2}{*}{YES} \\
 & 90 & 47.000 & 1.260 & $I\times$$I$ & & \\ \hline
\multirow{2}{*}{4} & 90 & 358.000 & 2.780 & $Q\times$$I$ &
\multirow{2}{*}{99.967} & \multirow{2}{*}{YES} \\
 & 180 & 271.000 & 1.790 & $I\times$$I$ & & \\ \hline
\multirow{2}{*}{5} & 90 & 73.667 & 3.480 & $Q\times$$I$ &
\multirow{2}{*}{94.667} & \multirow{2}{*}{YES} \\
 & 180 & 315.000 & 1.590 & $I\times$$I$ & & \\ \hline
\multirow{2}{*}{6} & 90 & 108.667 & 3.900 & $Q\times$$I$ &
\multirow{2}{*}{95.667} & \multirow{2}{*}{YES} \\
 & 180 & 196.333 & 1.460 & $I\times$$I$ & & \\ \hline
 \multirow{2}{*}{7} & 90 & 197.667 & 4.200 & $Q\times$$I$ &
\multirow{2}{*}{51.667} & \multirow{2}{*}{YES} \\
 & 90 & 172.333 & 2.100 & $I\times$$I$ & & \\ \hline
 \multirow{2}{*}{8} & 180 & 228.667 & 4.170 & $Q\times$$I$ &
\multirow{2}{*}{144.000} & \cellcolor{cyan} \\
 & 180 & 46.667 & 1.130 & $I\times$$I$ & & \multirow{-2}{*}{\cellcolor{cyan}NO} \\ \hline
\end{tabular}
\end{table}

As a result, if a maximum deviation of 40\textordmasculine\ over the quadrature condition ($\beta_{max} =$ 40\textordmasculine) is admitted, the phase detector can comply for the frequencies ranging from 3 GHz to 7 GHz. Regarding the input signal to the phase detector, the corrections performed in the calibration process of the generators, as well as those derived from the $S$ parameters of the connection block, equalize the power at the inputs of the phase detector and cause the phase difference between them to be determined by the phase difference reflected by the primary generator ($\theta_M$). However, the input amplitude is given by the power of the secondary generator ($P_S =$ 0 dBm) plus the effect of the $S_{a1}$ in dB for each frequency ($\Delta P$ in Table \ref{tab:table_iii}). In other words, $P_S$ +$\Delta P$ (dBm) would determine the input power of the phase detector.

\section{Conclusions}
A novel phase detector measurement procedure using quasi-synchronized RF generators that share a 10 MHz reference has been presented. This makes it possible to use two generators with lower cost and worse performance in terms of synchronization as compared to procedures with synchronized generators that have less use complexity but are much more limited in frequency range ($<$ 6 GHz).

The new procedure is characterized by continuously alternating calibration and measurement sequences to make up for the phase drift of quasi-synchronized generators. As the frequency increases, the procedure becomes more useful due to the output frequency difference between generators is greater and therefore the phase variation is faster.

The calibration and measurement system has been detailed in terms of equipment, devices, circuits, connections, sources of error, software, and methodology. The measurement stage comprises the acquisition of two sets of data: the reference voltages and the detector curves without phase reference obtained through the increase of the frequency of one of the generators. The procedure only requires a quick execution of the reference voltages acquisition in order to guarantee a maximum phase error specification ($\pm$1\textordmasculine\ in this paper but less is possible), which enables the use of a very simple control element based in Arduino Uno and cheaper instruments. Although time is limited to perform the measurement of the reference voltages, the semi-automatic procedure is suitable for the majority of phase detectors, with the exception of some topologies that require more time for phase determination \cite{Berlotti2010, Marcellis2015}.

The procedure has been applied to obtain the output voltage curves of a 360\textordmasculine\ phase detector base on switched dual multiplier from 3 GHz to 8 GHz at 1 GHz step. Moreover, it has been detailed step by step, which facilitates its reproducibility and the use of other equivalent instrumentation.

The errors arising by the signal distribution network have been measured in order to know how they should be treated in the measurement and calibration procedure, as well as to quantify their effect in case they had not been taken into account (Table \ref{tab:table_iii}). Not including this source of error would result in an increase of almost 1\textordmasculine\ (-0.94\textordmasculine\ at 4 GHz) over the initial specification ($\pm$1\textordmasculine) and 0.3 dB (8 GHz) in the power arriving at the phase detector, although that power variation usually has little influence over the phase detector response. Accordingly,  determining whether a VNA is necessary, the accuracy specifications of the  R\&S ZVK VNA is $\pm$1\textordmasculine\ and $\pm$0.1 dB between 2 GHz and 8 GHz \cite{1EZ48}. Therefore, the -0.94\textordmasculine\ is within the accuracy specifications and could obviate the use of the VNA, increasing the measurement uncertainty by $\pm$1\textordmasculine. That is, measurements have a final maximum error of $\pm$2\textordmasculine\ for both frequency and calibrated input power. Table \ref{tab:table_v} shows a general comparison between the method proposed in this paper and others detailed in the introduction.

\begin{table}[!t]
\caption{Comparison with others configurations: VL (Very Low), L, M (Medium), H (High), VH, and NA (Not Available).}
\label{tab:table_v}
\resizebox{0.5\textwidth}{!}
{
\begin{tabular}{|l|c|c|c|c|c|c|}
\hline
 \parbox{4.0cm}{\centering Measurement Procedures} &
 \rotatebox[origin=c]{90}{\parbox{1.4cm}{\centering Error}} &
 \rotatebox[origin=c]{90}{\parbox{1.4cm}{\centering VNA\\Mandatory}} &
 \rotatebox[origin=c]{90}{\parbox{1.4cm}{\centering Versatility}} &
 \rotatebox[origin=c]{90}{\parbox{1.4cm}{\centering Use\\Complexity}} &
 \rotatebox[origin=c]{90}{\parbox{1.4cm}{\centering Cost}} &
 \rotatebox[origin=c]{90}{\parbox{1.4cm}{\centering Frequency}}
\\ \hline
\rowcolor{color1}
1 Generator \& 2 RF Outputs \cite{Yang2014} & {$<\pm$0.1} &
No & VH & VL & VL & VL \\ \hline
\multicolumn{7}{|l|}{ 1 Generator \& 1 RF Output \&} \\ \hline
\quad PS (Power Splitter) \cite{Pogorzelski2005} &
NA & No & VL & VL & M & VH \\ \hline
\quad PS \& Multipliers \cite{Philippe2018} &
NA & No & M & L & VH & VH \\ \hline
\quad PS \& PhaseShift (Ph) \cite{Hua2016,Hua2011,Hua2009} &
NA & Yes & M & M & M & VH \\ \hline
\quad PS \& Ph \& Variable Gain \cite{Huang2022} &
NA & Yes & H & H & M & VH \\ \hline
\rowcolor{color2} \multicolumn{7}{|l|}{ 2 Generators} \\ \hline
\rowcolor{color2} \quad Synchronized \cite{Perez2021,1GP67,Umpierrez2012} &
$<\pm$0.1 & No & VH & H & VH & M \\ \hline
\rowcolor{color2} \quad This article & $<\pm$2\textordmasculine &
No & VH & VH & H & VH \\ \hline
\end{tabular}
}
\end{table}

Finally, the procedure used to obtain the correctly referenced $I\times$$I$ and $Q\times$$I$ curves that characterize the behavior of the phase detector has been described, and the phase difference between them has been calculated. Based on the maximum deviation on the quadrature condition \cite{Perez2021,Perez2017}, which has been arbitrarily set to a value of 40\textordmasculine\ (Fig. \ref{fig:figura5}), it is found that the phase detector will be able to operate in the frequency range of 3 GHz to 7 GHz (Table \ref{tab:table_iv}).

\section*{Acknowledgments}
Thanks to Juan Domingo Santana Urb\'in for his enormous contribution during this research. This work was supported by the Spanish Gobernment under Grant (PID2020-116569RB-C32) Project.

\bibliographystyle{IEEEtran}
\bibliography{mybibliography.bib}
 
\vspace*{-1.0cm}
\begin{IEEEbiography}[{\includegraphics[width=1in,height=1.25in,clip,keepaspectratio]{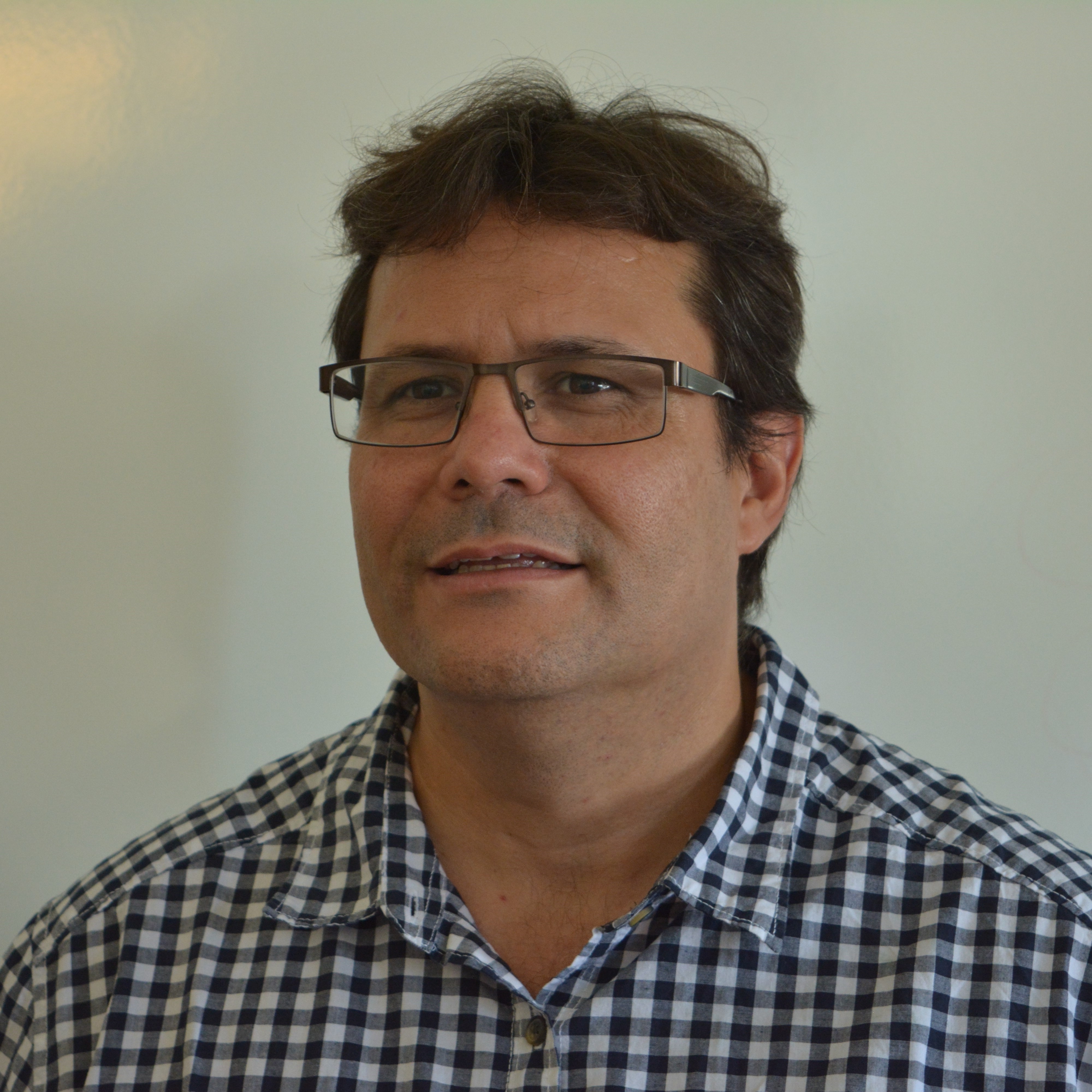}}]{V\'ictor Ara\~na-Pulido} (Member, IEEE) was born in Las Palmas, Spain, in 1965. He received the M.Sc. degree from the Universidad Polit\'ecnica de Madrid (UPM), Madrid, Spain, in 1990, and the Ph.D. degree from the Universidad de Las Palmas de Gran Canaria (ULPGC), Las Palmas, in 2004. He is currently an Associate Professor with the Signal and Communication Department and a member of the Institute for Technological Development and Innovation in Communications (IDeTIC), ULPGC. His current research interests include the nonlinear design of microwave circuits, control and communications subsystem units.
\end{IEEEbiography}
\vspace*{-1.0cm}
\begin{IEEEbiography}[{\includegraphics[width=1in,height=1.25in,clip,keepaspectratio]{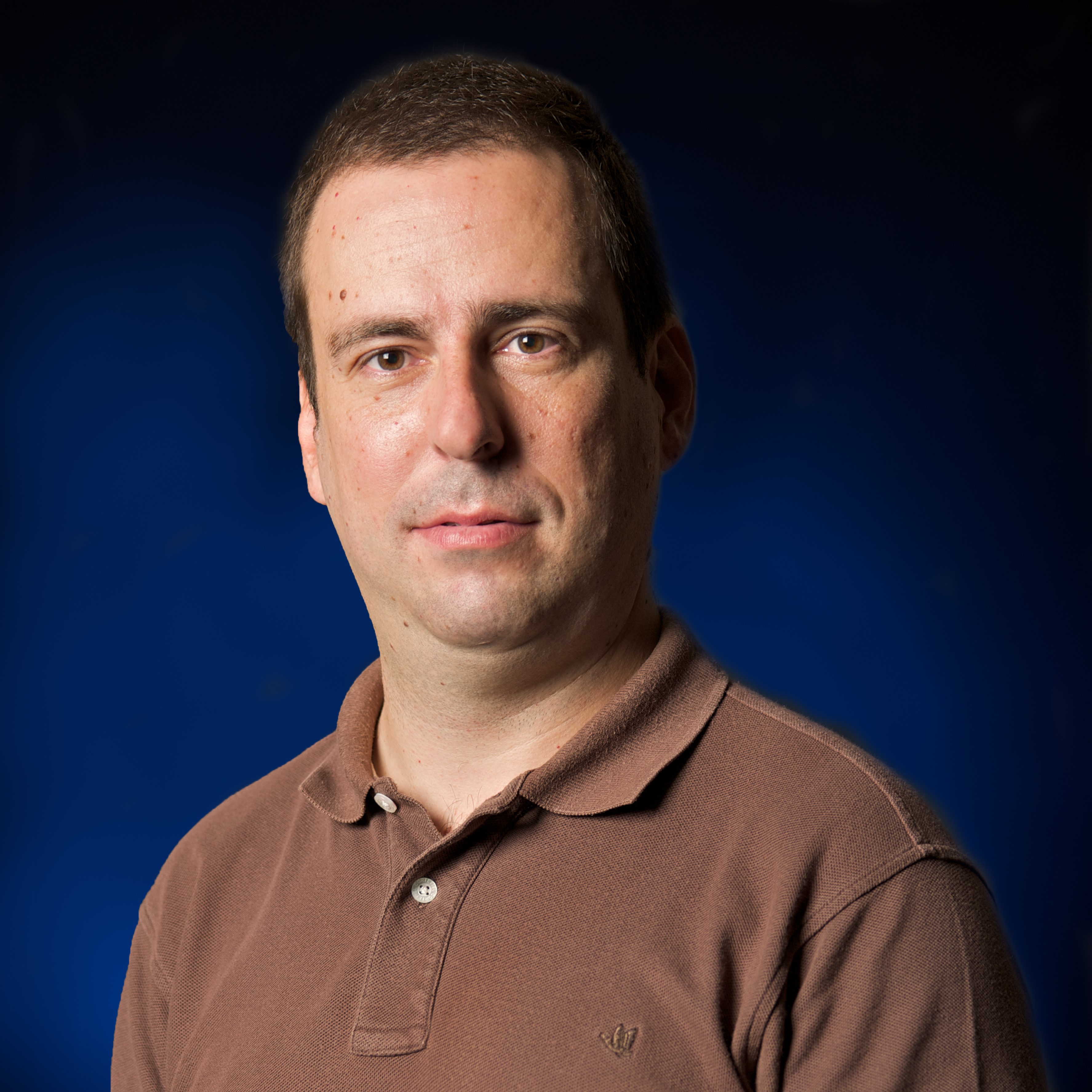}}]{Francisco Cabrera-Almeida}  (Member, IEEE) was born in Las Palmas, Spain, in 1970. He received the M.Sc. and Ph.D. degrees from the Universidad de Las Palmas de Gran Canaria (ULPGC), Las Palmas, in 1997 and 2012, respectively. He is currently an Assistant Professor with the Signal and Communications Department and a member of the Institute for Technological Development and Innovation in Communications (IDeTIC), ULPGC. His current research interests include numerical electromagnetic modeling techniques and radiowave propagation and communications systems applied to data acquisition complex networks.
\end{IEEEbiography}
\vspace*{-1.0cm}
\begin{IEEEbiography}[{\includegraphics[width=1in,height=1.25in,clip,keepaspectratio]{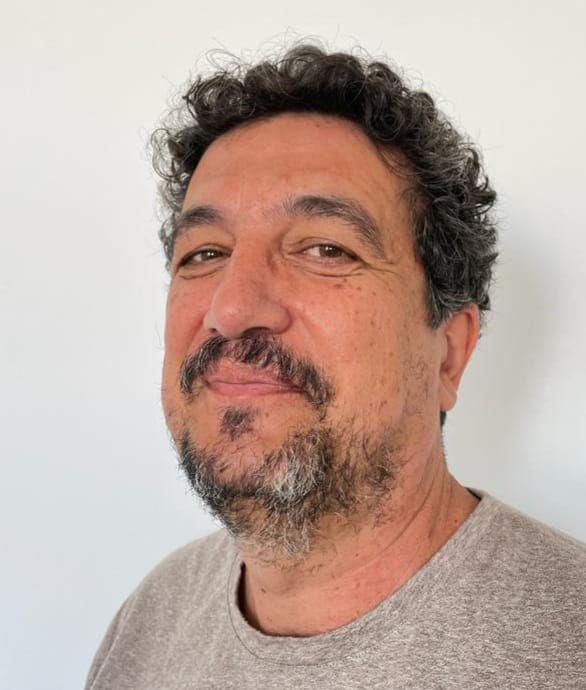}}]{Pedro Quintana-Morales}
was born in Santa Cruz de Tenerife, Spain, in 1964. He received the M.Sc. degree from the Universidad Politécnica de Madrid, (UPM) Madrid, Spain, in 1989, and the Ph.D. degree from the Universidad de Las Palmas de Gran Canaria (ULPGC), Las Palmas, Spain, in 2016. He is currently an Assistant Professor with the Signal and Communication Department and a member of Institute for Technological Development and Innovation in Communications (IDeTIC), ULPGC. His current research interests include signal processing, data analysis applied to speech, images, biosignals and sensor networks.
\end{IEEEbiography}
\vspace*{-1.0cm}
\begin{IEEEbiography}[{\includegraphics[width=1in,height=1.25in,clip,keepaspectratio]{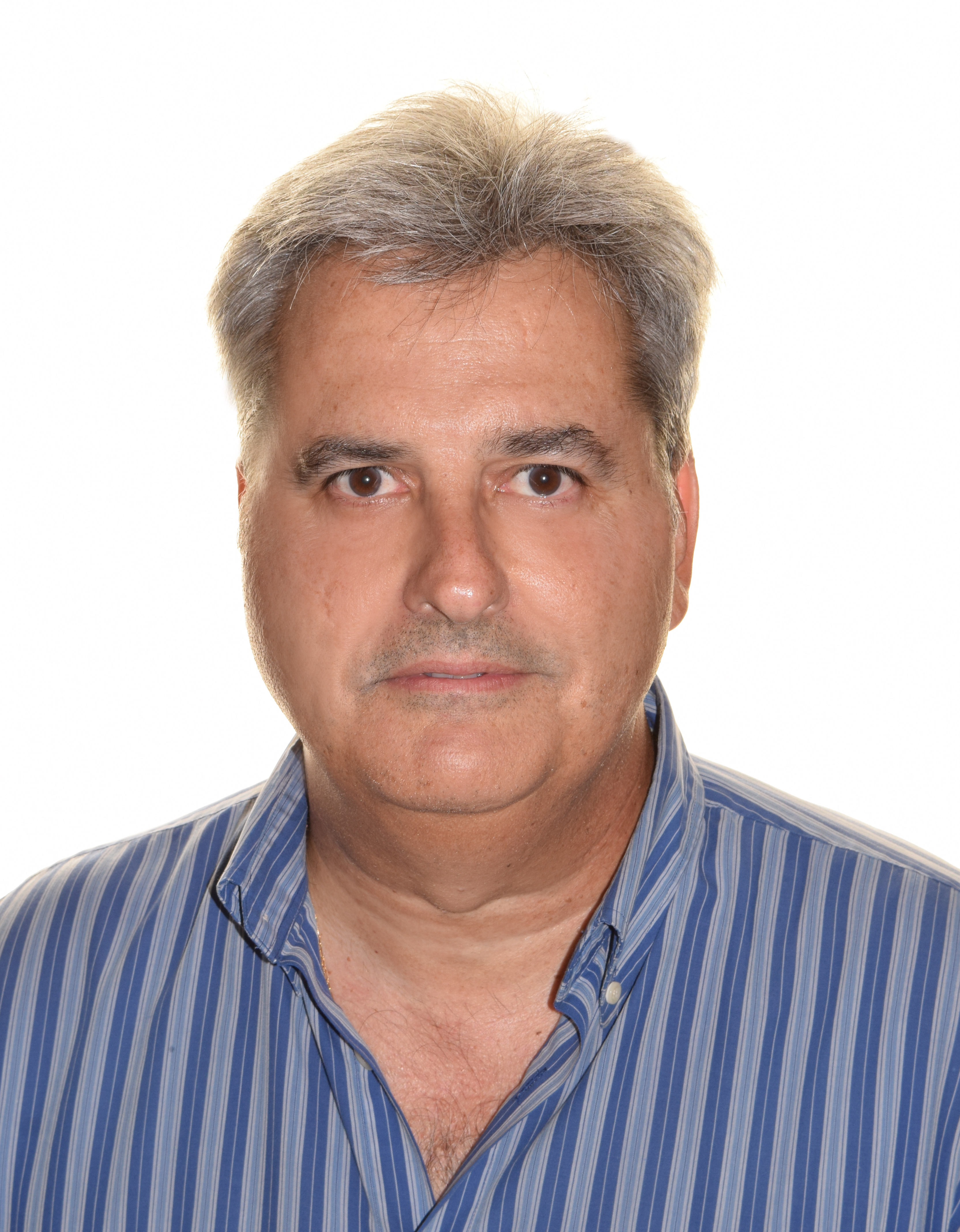}}]{Eduardo Mendieta-Otero} was born in La Habana, Cuba, in 1966. He received B.Sc. degrees in Electronic Equipment and Radio Communication from University of Las Palmas de Gran Canaria (ULPGC), Spain, in 1992 and 1994 respectively. He received the M.Sc. and Ph.D. degrees from ULPGC in 2001 and 2016, respectively. He is currently an Assistant Professor with the Signal and Communication Department and member of Institute for Technological Development and Innovation in Communications (IDeTIC), ULPGC. His research activities are mainly in the field of Signal Processing for Radio Communications.
\end{IEEEbiography}

\vfill

\end{document}